\documentstyle[12pt]{article}
\newcommand{\beq}{\begin{equation}}
\newcommand{\eeq}{\end{equation}}
\begin{document}
\begin{center}

{\bf Quasiparticle dispersion and lineshape in a strongly correlated
liquid with the fermion condensate}

\vspace{0.2cm}

M.Ya. Amusia$^{a,b}$, V.R. Shaginyan$^{c,}$ \footnote{E--mail:
vrshag@thd.pnpi.spb.ru}\\
\bigskip
$^{a\,}$The Racah Institute of Physics, Hebrew University,
Jerusalem 91904, Israel;\\ $^{b\,}$Physical-Technical Institute,
Russian Academy of Science, K-21 St. Petersburg, Russia;\\
$^{c\,}$ Petersburg Nuclear Physics Institute, Russian Academy of
Science, Gatchina, 188350 Russia\\

\end{center}
\begin{abstract}
A model of a strongly correlated electron liquid based on the
fermion condensation (FC) is applied to the consideration of high
temperature superconductors in its superconducting and normal states.
Within our model the appearance of FC presents a boundary, separating
the region of strongly interacting electron liquid from the region of
strongly correlated electron liquid. We show that at temperatures
$T\leq T_c$ the quasiparticle dispersion in systems with FC can be
presented by two straight lines, characterized by effective masses
$M^*_{FC}$ and $M^*_L$, respectively, and intersecting near the
binding energy which is of the order of the superconducting gap.
This same picture persists in the normal state being modified only by
the presence of the quasiparticle width. We argue that this strong
change of the quasiparticle dispersion can be enhanced in underdoped
samples due to strengthening the FC influence. The single
particle excitation width and lineshape are also studied.
\end{abstract}

\noindent {\it PACS:} 71.27.+a, 74.20.Fg, 74.25.Jb

\noindent {\it Keywords:} Strongly correlated electrons; Phase
transitions; Electronic structure

\bigskip\bigskip
Recent discovery of a new energy scale for quasiparticle dispersion
in Bi$_2$Sr$_2$CaCu$_2$O$_{8+\delta}$ in the superconducting
and in normal states \cite{blk,krc} have brought
new insight to the physics of high temperature superconductors (HTS),
imposing serious constraints upon the theory of HTS.
The newly discovered additional energy scale manifests itself as a
break in the quasiparticle dispersion near $(50-70)\,$meV, which
results in a drastic change of the quasiparticle velocity
\cite{blk,krc,vall}.  Such a behavior is definitely different from
what one could expect in a normal Fermi liquid.  Moreover,
this behavior can hardly be understood in the frame the marginal Fermi
liquids (MFL) theory, because central conclusions of MFL theory are
that the single-particle self energy has unimportant momentum
dependence and there exists the only relevant energy scale
which is defined by temperature $T$ of the normal state \cite{var}.

To describe a correlated liquid a conventional way can be used,
assuming that the correlated regime corresponds to the noninteracting
Fermi gas by adiabatic continuity in the same way
as in the framework of the Landau normal Fermi liquid theory.
A question is in order at this point: is it
possible?  Most likely the answer is negative. Thus, having in mind
to attack the above-mentioned problem we direct our attention to a
model in the frame of which a strongly correlated liquid is separated
from conventional Fermi liquid by a phase transition related to the
onset of FC \cite{ks,vol}. The aim of our Letter is to show that
without any adjustable parameters the energy scale for quasiparticle
dispersion is naturally explained within the presented model.  Our
Letter is organized as follows. First of all we focus on the general
features of Fermi systems with FC. After that, we describe the
quasiparticle dispersion and lineshape. Finally, we summarize the
main results of our Latter.

Let us start by considering the key points of the FC theory.
FC is related to a new class of solutions of
the Fermi-liquid-theory equation \cite{lan},
\beq \frac{\delta (F-\mu N)}{\delta
n(p,T)}=\varepsilon(p,T)-\mu(T)-T\ln\frac{1-n(p,T)}{n(p,T)}=0,\eeq
for the quasiparticle distribution function $n(p,T)$, which depends
on the momentum $p$ and temperature $T$. Here $F$ is the free energy,
and $\mu$ is the chemical potential, while,
\beq\varepsilon(p,T)=\frac{\delta E[n(p)]}{\delta n(p,T)},\eeq
is the quasiparticle energy, being a functional of
$n(p,T)$ just like the energy $E[n(p)]$ and the other thermodynamic
functions. Eq. (1) is usually presented in the
form of the Fermi-Dirac distribution, \beq
n(p,T)=\left\{1+\exp\left[\frac{(\varepsilon(p,T)-\mu)}
{T}\right]\right\}^{-1}.\eeq In homogeneous matter and at $T\to0$,
one gets from Eqs. (1),(3) the standard solution
$n_F(p,T\to0)\to\theta(p_F-p)$,
with $\varepsilon(p\simeq p_F)-\mu=p_F(p-p_F)/M^*_L$, where $p_F$ is
the Fermi momentum, and $M^*_L$ is the generally used effective mass
\cite{lan},
\beq \frac{1}{M^*_L}
=\frac{1}{p}\frac{d\varepsilon(p,T=0)}{dp}|_{p=p_F}.\eeq
It is implied that $M^*_L$ is positive and finite at the Fermi
momentum $p_F$. As a result, the $T$-dependent corrections to
$M^*_L$, to the quasiparticle energy $\varepsilon (p)$, and other
quantities, start with $T^2$-terms.
But this solution is not the only one possible. There
exist ``anomalous" solutions of Eq. (1) associated
with the so-called fermion condensation \cite{ks,ksk,dkss}. Being
continuous and satisfying the inequality $0<n(p)<1$ within some
region in $p$, such solutions $n(p)$ admit a finite limit for the
logarithm in Eq. (1) at $T\rightarrow 0$ yielding,
\beq \varepsilon(p)=\frac{\delta E[n(p)]}
{\delta n(p)} =\mu, \: p_i\leq p \leq p_f. \eeq
Equitation (5) leads to the minimal
value of $E$ as a functional of $n(p)$ when in system under
consideration a strong rearrangement of the single particle
spectra can take place. We see from Eq. (5) that the occupation
numbers $n(p)$ become variational parameters: the solution $n(p)$
takes place if the energy $E$ can be lowered by alteration of the
occupation numbers. Thus, within the region $p_i<p<p_f$, the
solution $n(p)$ deviates from the Fermi step function $n_F(p)$ in
such a way that the energy $\varepsilon(p)$ stays constant while
outside this region $n(p)$ coincides with $n_F(p)$. As a result, the
standard Kohn-Sham scheme for the single particle equations is no
longer valid beyond the point of the FC phase transition \cite{vsl}.
Such a behavior of systems with FC is clearly different from what one
expects from the well known local density calculations. Therefore
these calculations are hardly applicable to describe systems with FC.
As to the quasiparticle formalism, it is applicable to this problem
since, as it will be demonstrated, the damping of single particle
excitations is not large compared to their energy \cite{dkss}. It is
also seen from Eq. (5) that a system with FC has the well-defined
Fermi surface.
The new solutions, according to Eq. (1),
within the interval occupied by the fermion condensate
have at low $T$ the shape of the spectrum $\varepsilon(p,T)$ linear
in T \cite{dkss,kcs},  \beq
\varepsilon(p,T)-\mu(T)=T\ln\frac{1-n(p)}{1-(1-n(p))}
\simeq T(1-2n(p)) \sim T<<T_f. \eeq
We note that at $T\ll T_f$ the occupation numbers $n(p)$ are
approximately independent of $T$, being given by Eq. (5).
Here $T_f$ is the quasi-FC phase transition
temperature above which FC effects become insignificant \cite{dkss},
\beq
\frac{T_f}{\varepsilon_F}\sim\frac{p_f^2-p_i^2}{2M\varepsilon_F}
\sim\frac{\Omega_{FC}}{\Omega_F},\eeq
where $\Omega_{FC}$ stands for the condensate volume, $\varepsilon_F$
for the Fermi energy, and $\Omega_F$ for the volume of the Fermi
sphere.
One can imagine that at $T>0$ dispersionless plateau
$\varepsilon(p)=\mu$ given by Eq. (5) is slightly turned
counter-clockwise about $\mu$.  As a result, the plateau is just a
little tilted and rounded off at the end points.
If $T\ll T_f$, then according to Eqs. (1) and (6) the effective mass
$M^*_{FC}$ related to FC is given by,
\beq \frac{p_F}{M^*_{FC}}\simeq \frac{2T}{p_f-p_i}.\eeq
To obtain Eq. (8) an approximation for the derivative
$dn(p)/dp\simeq -1/(p_f-p_i)$ is used. Then, having in mind that
$p_f-p_i\ll p_F$, the following estimates for the effective
mass $M^*_{FC}$ are obtained,
\beq \frac{M^*_{FC}}{M} \sim
\frac{N(0)}{N_0(0)}\sim\frac{T_f}{T},\eeq
that show the temperature
dependence of the effective mass. Here $M$ denotes the bare
electron mass, $N_0(0)$ is the density of states of
noninteracting electron gas, and $N(0)$ is the density of states
at the Fermi level.
Multiplying both sides of Eq. (8) by $p_f-p_i$ we obtain the energy
scale $E_0$ separating the slow dispersing low energy part,
related to the effective mass $M^*_{FC}$, from the faster dispersing
relatively high energy part, defined
by the effective mass $M^*_{L}$,
\beq E_0\simeq 2T.\eeq
It is seen from Eq. (10) that the scale $E_0$ does not depend on the
condensate volume. Consider HTS compounds, then it is suggested that
FC arises around the van Hove singularities, causing as it follows
from Eqs. (8) and (9), the large density of states and the large
value of the difference $p_f-p_i$ at the singularities. Then, the
volume and difference depend on the point of the Fermi surface, say,
on the angle $\phi$ along the Fermi surface, nonetheless, the
magnitude $E_0$ remains constant, being independent of the angle.
This differs essentially from the case for the effective mass
$M^*_{FC}$, that can be strongly dependent upon the angle through the
difference $p_f(\phi)-p_i(\phi)$, as it is seen from Eq. (8). It is
pertinent to note that outside the FC region the single particle
spectrum is insignificantly affected by the temperature, being defined
by the ordinary effective Landau mass $M^*_L$, given by Eq. (4),
however calculated at $p\leq p_i$ instead of at $p=p_F$. Thus, we
come to the conclusion that a system with FC is
characterized by two effective masses:  $M^*_{FC}$ that is related to
the single particle spectrum at low energy scale, and $M^*_L$
dealing with the spectrum at higher energy scale.  These two
effective masses manifest itself as a break in the quasiparticle
dispersion, which can be approximated by two straight lines
intersecting at the energy $E_0$.  This break is observed at
temperatures $T\ll T_f$, and, as we will see below, at $T\leq T_c$,
when the superconducting state is grounded on the FC state. In
this case the occupation numbers in the momentum area occupied by the
fermion condensate are slightly disturbed by the
superconducting pairing correlations so that effective mass
$M^*_{FC}$ becomes large but finite even at $T=0$. Note that at
comparatively low temperatures FC and the superconductivity go
together due to remarkable peculiarity of the FC phase
transition. Indeed, this transition is related to spontaneous
breaking of gauge symmetry: at $T=0$ the superconductivity order
parameter $\kappa(p)=\sqrt{n(p)(1-n(p))}$ has a nonzero value over
the region occupied by the fermion condensate, while the gap $\Delta$
may vanish \cite{dkss,vsl}.

It is seen from Eq. (5) that at the point of FC phase transition
$p_f\to p_i\to p_F$, while the effective mass and the density of
states, as it follows from Eqs. (5),(9), tend to infinity. One can
conclude that at $T=0$ the beginning of the FC phase transition is
connected to the absolute growth of $M^*_{FC}$. The onset of the
charge density-wave instability in a many electron system, such
as electron liquid, which takes place as soon as the effective
inter-electron constant reaches its critical value $r_{cdw}$, is to
be preceded by the unlimited growth of the effective mass \cite{ksz}.
In a simple liquid, the effective constant turns out to be
proportional to the dimensionless average distance $r_s\sim r_0/a_B$
between particles (electrons or holes) of system in question, with
$r_0$ being the average distance, and $a_B$ is the Bohr radius.
The physical reason for this growth is the
contribution of the virtual charge density fluctuations to the
effective mass. Note that the excitation energy of these
fluctuations become very small if $r_s\simeq r_{cdw}$.  Thus, when
$r_s\sim r_{cdw}$, FC can take place. So, the standard Fermi liquid
behavior can be strongly disturbed and completely broken by strong
charge fluctuations as soon as the insulator regime is approached in
a continuous fashion. Let us remind that the charge-density wave
instability takes place in three-dimensional (3D) \cite{lp} and
two-dimensional (2D) electron liquids \cite{sns} at sufficiently
high $r_s$. As soon as $r_s$ reaches its critical value the FC phase
transition takes place. Thereafter, the condensate volume turns out
to be proportional to $r_s$ as well as $T_f/\varepsilon_F\sim r_s$
\cite{dkss,ksz}. Thus, we can accept a simple model relating systems
with FC to strongly correlated metals: we suggest that the effective
coupling constant $r_s$ increases with decreasing doping. This
proposal assumes that both quantities $T_f$ and condensate volume
$\Omega_{FC}$ are increasing with decreasing doping. Thus, these
are higher in underdoped samples comparatively to overdoped
ones. Then, FC serves as a stimulating source of new phase
transitions which lift the degeneracy of the spectrum.  For example,
FC can generate the spin density wave phase transition, or
antiferromagnetic one, thus leading to a variety of the system's
properties. The superconducting phase transition is also aided by FC.
The simple consideration presented above explains extremely large
variety of HTS properties.  In order to analyze the quasiparticle
dispersion and lineshape at $T\leq T_c$, we are going to investigate
the situation when the superconductivity wins the competition with
the other phase transition up to the critical temperature $T_c$.

We now discuss the origin of two effective masses
$M^*_L$ and $M^*_{FC}$ in the superconducting state resulting in the
nontrivial quasiparticle dispersion and in a change of the
quasiparticle velocity. As we will see below our results are in a
reasonably good agreement with experimental data
\cite{blk,krc,vall}. To simplify the discussion
let us put $T=0$.
The ground state energy $E_{gs}$ of a system in the superconducting
state is given by the formula,
\beq E_{gs}[\kappa({\bf p})]=E[n({\bf p})]+
E_{sc}[\kappa({\bf p})], \eeq
where the occupation numbers $n({\bf p})$ are connected
to the order parameter
$\kappa({\bf p})=\sqrt{n({\bf p})(1-n({\bf p}))}$.
The second term $E_{sc}[\kappa_p]$ on the right hand side of Eq. (11)
is defined by the superconducting contribution which in the simplest
case is of the form,
\beq E_{sc}[\kappa_p]=\int V_{pp}({\bf p}_1,
{\bf p}_2)\kappa({\bf p}_1) \kappa({\bf p}_2)
\frac{d{\bf p}_1d{\bf p}_2}{(2\pi)^4}. \eeq
Consider a 2D liquid on a simple square
lattice which is in the superconducting state with d-wave symmetry of
the order parameter $\kappa({\bf p})$. In such a case, the long-range
component in momentum space of particle-particle interaction $V_{pp}$
is repulsive. The sort-range component is relatively dominant and
attractive at the small momenta \cite{abr}. Then, we suppose that FC
arises near the Van Hove singularities, causing, as it follows from
Eq. (9), large density of states at these points \cite{kcs}. Under
such circumstances, as a first approximation we assume the different
regions of the maximal value $\Delta_1$ of the gap $\Delta$ are
disconnected \cite{abr,vs,ams}. Therefore, we can approximate the
particle-particle interaction by its short-range component, it is
assuming that $V_{pp}(q)\simeq V_2\delta(q)$. Varying $E_{gs}$ given
by Eq. (11) with respect to $\kappa({\bf p})$ one finds, \beq
(\varepsilon({\bf p})-\mu)-\Delta({\bf p})\frac{1-2n({\bf p})}
{2\sqrt{n({\bf p})(1-n({\bf p}))}}=0.  \eeq Here $\varepsilon({\bf
p})$ is defined by Eq. (2), and,
\beq \Delta({\bf p})=-\int V_{pp}
({\bf p},{\bf p}_1) \sqrt{n({\bf p}_1)(1-n({\bf p}_1))}
\frac{d{\bf p}_1}{4\pi^2}.  \eeq
A few remarks are in order at this point.  If
$V_2\to 0$, then $\Delta({\bf p})\to 0$, and Eq. (13) reduces to,
\beq
\varepsilon({\bf p})-\mu=0,\,\,\,\, {\mathrm {if}}\,\,\,
0<n({\bf p})<1;\,\,\, \kappa({\bf p})\neq 0,
\eeq
producing FC solutions, defined by Eq. (5) \cite{dkss,vsl}.
Taking into account Eq. (15), we come to the conclusion that
if $V_2$ is sufficiently small, then in the first order in $V_2$
the function $\kappa({\bf p})$ is defined by Eqs. (5),(15).
The gap $\Delta$ is given by Eq. (14), while $n({\bf p})$ is {\it
fixed} by Eq. (5), and as it is seen the gap is linear in the
coupling constant of the particle-particle interaction \cite{ks}.

As soon as coupling constant $V_2$ becomes finite but small,
plateau $\varepsilon({\bf p})-\mu=0$ is slightly
tilted and rounded off at the end points. As a result,
taking into account the $\delta$-like shape of the particle-particle
interaction, one gets from Eqs. (13) and (14),
\beq \varepsilon({\bf
p})-\mu\simeq -\frac{V_2}{2}(1-2n({\bf p})), \eeq
which allows to estimate the
effective mass,
\beq \frac{p_F}{M^*_{FC}}\simeq-\frac{V_2}{p_f-p_i}
\simeq\frac{2\Delta_1}{p_f-p_i}.
\eeq
Now, one can directly conclude from Eq. (17),
\beq E_0\simeq\frac{(p_f-p_i)p_F}{M^*_{FC}}\simeq 2\Delta_1.
\eeq
Note that calculations in the frame of a simple model land
additional support to the presented consideration \cite{dkss}.
Eq. (17) allows to estimate effective mass $M^*_{FC}$  connected to
the region occupied by FC at temperatures $T\leq T_c$. Outside
this region the effective mass is defined by $M^*_L$, being roughly
independent of the pairing correlations and temperature. It is seen
from Eq. (18), that again this time at $T\leq T_c$, the
quasiparticle dispersion can be presented by two straight lines
characterized by two effective masses, $M^*_{FC}$ and $M^*_L$,
respectively, and intersecting near binding energy
$E_0\simeq2\Delta_1$. The break
separating the faster dispersing high energy part, related to mass
$M^*_L$ from the slower dispersing low energy part defined by
$M^*_{FC}$, is likely to be more pronounced in underdoped samples at
least because of the rise of temperature $T_f$, since the lower is
the doping, the higher is $T_f$. We remind that in accordance with
our assumption the condensate volume $\Omega_{FC}$ is growing with
the underdoping. It was also suggested that FC arises
near the van Hove singularities, while the different areas of FC
overlap only slightly. Therefore, it follows from Eq. (17), that as
one moves from the node of the gap towards its maximal value
$\Delta_1$, that is from minimal value of $\Omega_{FC}(\phi)$ towards
the maximal one, the ratio $M^*_{FC}/M^*_{L}$ grows in magnitude,
transforming the dispersion kink into a distinct break. In fact,
suggesting that temperature $T_f$ depends on the angle $\phi$ along
the Fermi surface, that is $T_f(\phi)$, and taking into account that
$T_f$ reaches its maximal value at the fermion condensate area, one
can arrive at the same conclusion. The dispersions above $T_c$
exhibit the same structure except that effective mass $M^*_{FC}$ is
governed by Eq. (8) rather then (17), and both the dispersion and
break are disguised partly  by the quasiparticle width.  Thus, one
can conclude that there also exists a new energy scale at $T\ll T_f$
defined by $E_0$ and closely related to $T_f$ \cite{ars}.

Now let us turn to the quasiparticle excitations with
the energy $E(\phi)=\sqrt{\varepsilon^2(\phi)+\Delta^2(\phi)}$.
At temperatures $T<T_c$ these are typical excitations of the
superconducting state. Here we are going to perform a qualitative
treatment of processes contributing to the width $\gamma$.  Within
the frame of this analysis, we can take gap $\Delta\simeq 0$,
considering excitations at the node of the gap. Then, our
treatment holds for both $T\leq T_c$ and $T_c\leq T$. For
definiteness, we consider the decay of a particle implying that
momentum $p>p_F$. Then, the width $\gamma(p,\omega)$ is given by
\cite{rit,kss},
\beq \gamma(p,\omega)=2\pi\int
\left|\frac{V(q)}{\epsilon(q,-\omega_{pq})}\right|^2
n({\bf k})(1-n({\bf k+q}))\delta(\omega_{pq}+\omega_{kq})
\frac{d{\bf q}d{\bf k}}{(2\pi)^4},
\eeq
with $\epsilon(q,-\omega_{pq})$ being the
complex dielectric constant, and $V(q)/\epsilon$ is
the effective interparticle interaction. Here ${\bf q}$ and
$\omega_{kq}=\varepsilon({\bf k+q})-\varepsilon({\bf k})$
are the transferred momentum and energy, respectively, and
$\omega_{pq}=\omega-\varepsilon({\bf p-q})$
is the decrease in the energy
of the quasiparticle as a result of the rescattering processes:
quasiparticle with energy $\omega$ decays into
two quasiparticles $\varepsilon({\bf p-q})$
and $\varepsilon({\bf k+q})$ and quasihole $\varepsilon({\bf k})$.
The transferred momentum $q$ must satisfy the condition,
\beq p>|{\bf p-q}|>p_F.\eeq
Eq. (19) gives the width as a function of $p$ and $\omega$,
while the width of a quasiparticle of
energy $\varepsilon(p)$ is given by
$\gamma(p,\omega=\varepsilon(p))$.
Estimating the width given by Eq. (19)
with the limitation (20) and $\omega_{pq}\sim T$,
one finds  for normal Fermi liquids
\beq\gamma(p,\omega=\varepsilon(p))\sim
\left(M^*_{L}\right)^3T^2.\eeq
In the case of FC one could estimate $\gamma\sim 1/T$
upon using Eqs. (9) and (21) \cite{noz}. This estimate were correct if
the dielectric constant is small, but it is not the case since
$\epsilon \sim M^*_{FC}$.
Now in the event of FC one has,
\beq
\gamma(p,\omega=\varepsilon(p))\sim
\frac{\left(M^*_{FC}\right)^3T^2}{\left(M^*_{FC}\right)^2}
\sim T \frac{T_f}{\varepsilon_F},\eeq
here $\varepsilon_F$ is the
Fermi energy \cite{dkss,kss}. Calculating $\gamma(p,\omega)$ as a
function of $p$ at a constant value of $\omega$, the same
result is obtained for the width that is given by Eq. (22) when
$\omega=\varepsilon(p)$.  The calculated function can be fitted with
a simple Lorentzian, since quasiparticles and quasiholes engaged
in the decay process are also located in the vicinity of the Fermi
level provided $\omega-\varepsilon_F\sim T$.  Thus,
the well-defined excitations exist at the Fermi
surface even in the normal state \cite{dkss,kss}. This result is in
accord with the experimental findings determined from the scans at
constant binding energies, (momentum distribution curves or MDCs)
\cite{vall,vall1}. On the other hand, considering $\gamma(p,\omega)$
as a function of $\omega$ at constant $p$, one can check that the
quasiparticles and quasiholes contributing to $\gamma(p,\omega)$ may
have the energy of the same order of the magnitude. Then, when
$\omega-\varepsilon_F\sim T$, the function is of the same Lorentzian
form, otherwise the shape of the function at high $\omega$ is
disturbed by high energy excitations that hardly can be considered as
well defined quasiparticles or quasiholes. Then the special form of
the quasiparticle dispersion, characterized by the two effective
masses, should be taken into consideration. In view of the fact that
the contribution of these excitations related to effective mass
$M^*_L$ is enhanced by the presence of FC, and
taking into account the background at high $\omega$, one can conclude
that the peak inevitably has a broadening which can hardly be
interpreted as standard width defined by either Eqs. (21) or (22). On
the other hand, one may follow the procedure suggested in \cite{krc},
using the Kramers-Kr\"{o}nig transformation to construct the
imaginary part of the self-energy starting with the real one.  As a
result, the lineshape of the quasiparticle peak as a function of
$\omega$ possesses a complex peak-dip-hump
structure directly defined by the existence of two effective masses
$M^*_{FC}$ and $M^*_L$, that give rise to the quasiparticle
dispersion approximated by two straight lines.
Under these conditions the lineshape is of
non-Lorentzian, which coincides with observational data obtained
from constant momentum scans (EDCs) \cite{blk,krc,kmf}. It is worth
to note that our consideration shows that it is the spectral peak
obtained from MDCs that provides important information on the
existence of well-defined excitations at the Fermi level and their
width \cite{ars}. The detailed numerical results will be presented
elsewhere.

In summary, we have discussed a model of a strongly correlated
electrons which has the FC phase transition and applied it to a
consideration of high temperature superconductors. The FC transition
fulfills the role of a boundary, separating the regions of strongly
interacting electrons and strongly correlated
electrons. At $T\leq T_c$ the quasiparticle dispersion in
systems with FC can be represented by two straight lines,
characterized by effective masses $M^*_{FC}$ and $M^*_L$,
respectively, and intersecting near binding energy $E_0\sim\Delta_1$.
We have also shown that this effect persists at $T_f\gg T\geq T_c$.
It is argued that this strong change of the quasiparticle dispersion
at $E_0$ can be enhanced in underdoped samples because of
strengthening the FC influence. The single particle excitations and
their width $\gamma$ are also studied.  It is demonstrated that
well-defined excitations with $\gamma\sim T$ exist at the Fermi level
even in the normal state. This result is in line with the
experimental findings derived from scans at constant binding
energies. As to the dispersion of the quasiparticle peak
obtained from constant momentum scans, it has a complex
peak-dip-hump structure directly related to the existence of two
effective masses $M^*_{FC}$ and $M^*_L$, that determine the
quasiparticle velocity showing a break near the binding energy $E_1$.

This research was funded in part by the Russian Foundation for
Basic Research under Grant No. 98-02-16170.

\end{document}